\documentstyle[11pt,newpasp,twoside,epsf]{article}
\begin{document}
\title{The Cosmic Background Imager}
\author{T. J. Pearson, B. S. Mason, S. Padin, A. C. S. Readhead,\\
 M. C. Shepherd, J. Sievers, P. S. Udomprasert, and J. K. Cartwright}
\affil{California Institute of Technology, Pasadena, California 91125, USA}

\begin{abstract}
The Cosmic Background Imager (CBI) is an instrument designed to make
images of the cosmic microwave background radiation and to measure its
statistical properties on angular scales from about 3 arc minutes to one
degree (spherical harmonic scales from $l \sim 4250$ down to $l \sim
400$).  The CBI is a 13-element interferometer mounted on a 6 meter
platform operating in ten 1-GHz frequency bands from 26 GHz to 36
GHz. The instantaneous field of view of the instrument is 45 arcmin (FWHM)
and its resolution ranges from 3 to 10 arcmin; larger fields can be
imaged by mosaicing. At this frequency and resolution, the primary
foreground is due to discrete extragalactic sources, which are
monitored at the Owens Valley Radio Observatory and subtracted from
the CBI visibility measurements.

The instrument has been making observations since late 1999 of both
primordial CMB fluctuations and the Sunyaev-Zeldovich effect in
clusters of galaxies from its site at an altitude of 5080 meters near
San Pedro de Atacama, in northern Chile. Observations will continue
until August 2001 or later.  We present preliminary results from the
first few months of observations.
\end{abstract}

\section{Introduction}

The Cosmic Background Imager (CBI) is a radio interferometer with
thirteen 0.9-meter diameter antennas mounted on a 6-meter tracking
platform. It operates in ten 1-GHz frequency channels from 26 to 36
GHz. Like several other instruments described at this Symposium, it is
dedicated primarily to the measurement of the angular power spectrum $C_l$
of the cosmic microwave background radiation (CMBR).  The CBI is
sensitive to multipoles in the range $400 < l < 4250$. The low end of
this range is also covered by the recent results from Boomerang (de
Bernardis et al.\ 2000; Lange et al.\ 2000) and Maxima (Hanany et al.\
2000; Balbi et al.\ 2000), so the CBI will be able to check those
results and extend the spectrum to higher $l$. The Degree Angular
Scale Interferometer (DASI), described by Erik Leitch at this
Symposium, is a lower-resolution sister project to the CBI, covering
$140 < l < 900$. The DASI and the CBI share many design elements, and the
telescope control software and the CBI correlator and receiver control
electronics have been duplicated by the DASI team for use with the
DASI project.

In addition to measuring the CMBR power spectrum, the CBI is being
used to image the Sunyaev-Zeldovich effect in clusters of galaxies, to
detect or place limits on the polarization of the CMBR, and to
characterize foreground emission in the 26--36 GHz band. We have also
used it to image supernova remnants, H\thinspace{\sc ii} regions, and
other discrete sources.

As the techniques involved in interferometric observations of the CMBR
are very different from those used in total-power and switched-beam
experiments, interferometers are a valuable alternative to such
experiments.  A single interferometer baseline is sensitive to a
narrow range of spatial frequencies, and so the square of the measured
visibility provides a direct estimate of the power spectrum in a
corresponding range of $l$. The center of this range is $l \approx
2\pi u$, where $u$ is the length of the interferometer baseline in
wavelengths, and the width of the range is set by the size of the
antenna aperture, with smaller antennas giving higher resolution in
$l$. The instantaneous field of view (``primary beam'') of the
interferometer is inversely proportional to the size of the antenna.
Thus by adjusting the baseline lengths and the size of the apertures
it is possible to match the characteristics of the instrument to the
scientific goals.  Two major advantages of interferometers are that
they are insensitive to extended emission, including the isotropic
background and much of the atmospheric emission, and that most
systematics (receiver gain instabilities, for example) are
uncorrelated between antennas and thus do not affect the measurements
significantly. Interferometers make images of the sky emission
convolved with an accurately known, adjustable point-spread function.

\section{Instrument Design}

The CBI is shown in Figure~1, and its major characteristics are shown
in Table~1.  More details of the design may be found in three papers
by Padin et al.\ (2000a, b, d), and on the CBI web page
(http://www.astro.caltech.edu/$\sim$tjp/CBI/).

\begin{figure}
\plotone{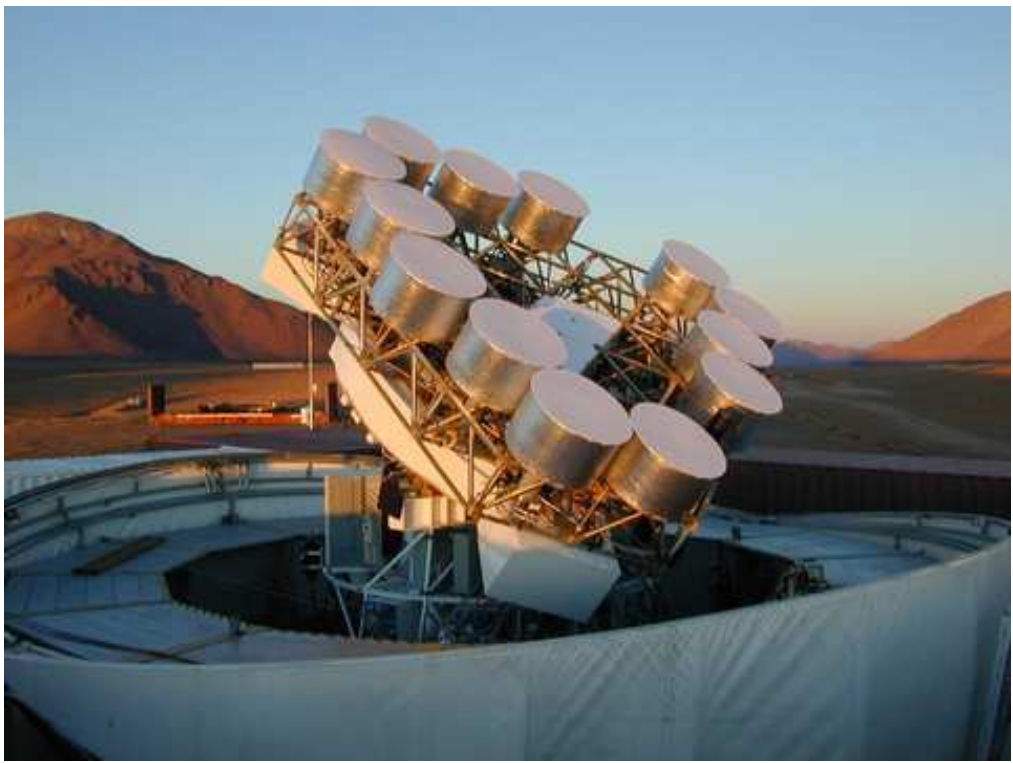}
\caption{View of the CBI on the Llano de Chajnantor, with the antennas
in a ring configuration.}
\end{figure}

The CBI operates in the frequency band 26--36~GHz, which was chosen as a
compromise between the effects of astronomical foregrounds and
atmospheric emission and the sensitivity that can be achieved with
HEMT amplifiers.

\begin{table}
\caption{CBI Specifications}
\begin{tabular}{ll}
\tableline
Observing frequency: & 26--36 GHz (wavelength $\sim1$~cm) \\
Number of channels:  & 10 channels, each 1~GHz wide \\
Number of antennas:  & 13 \\
Number of baselines: & 78 \\
Receivers:           & HEMT amplifiers, cooled to 6 K \\
Correlator:          & 780 analog complex correlators \\
Antenna:             & Cassegrain, 0.90 m diameter \\
Primary beam:        & Gaussian FWHM 45 arcmin \\
Minimum baseline:    & 1.00 m ($l \sim 630$) \\
Maximum baseline:    & 5.51 m ($l \sim 3500$) \\
Synthesized beamwidth: & FWHM 3--10 arcmin  \\
System temperature:  &  25 K \\
Noise in visibility measurements: & 4.5 Jy s$^{1/2}$ rms in each 1-GHz channel \\
Expected signal ($T_{\rm rms} = 40\,\mu$K): & $\sim 44$ mJy rms on 1-m baseline \\
\tableline
\end{tabular}
\end{table}

The key design challenges in the project were eliminating cross-talk
in a compact array and developing a wide-band correlator.  Receiver
noise scattering between adjacent antennas (cross-talk) causes false
signals at the correlator output and this could limit the sensitivity of
the instrument. We developed a shielded Cassegrain antenna with low
scattering to reduce the cross-talk (Padin et al.\ 2000a). The antennas
have machined, cast aluminum primaries which sit at the bottom of deep
cylindrical shields. The upper rims of the shields are rolled with a
radius of a few wavelengths to reduce scattering from the shield
itself. The secondaries are made of carbon fiber epoxy, to minimize
weight, and supported on transparent polystyrene feed legs. Cross-talk
between the antennas is $<-110$ dB in any CBI band.

The antennas are mounted on a rigid tracking platform 
supported by an altazimuth mount that is fully steerable to elevations
$>42\deg$. The antenna platform can be rotated about the optical
axis. In normal observations, the platform tracks the parallactic
angle so that observations are made at fixed $(u,v)$ points: i.e., the
baseline orientations are fixed relative to the sky.  Additional
discrete steps in the orientation of the platform are used to change
the baseline orientations and thus sample more $(u,v)$ points.

All the electronics, including the local oscillator systems and the
correlators, are also mounted on the antenna platform. Because the
antennas do not move relative to one another, all the receivers have
equal, stable paths to the correlator and no delay tracking is
required. The correlator (Padin et al.\ 2000b) is an analog filter bank
correlator with ten 1-GHz bands.  The design is based on a hybrid
module that contains all the signal distribution and multipliers for a
1-GHz band for 13 antennas. The module has a square grid of microstrip
transmission lines which distribute signals from the antennas to
Gilbert cell multiplier chips and tunnel-diode total power
detectors. The module is made entirely of chip components, with
wire-bond connections, and measures just 100 $\times$ 70 mm. A fast
phase-switching scheme, in which the receiver local oscillators are
inverted in Walsh function cycles, is used to reject cross-talk and
low-frequency pickup in the signal processing system.  Quadrature
errors in the correlator are measured by injecting correlated wide-band
noise just ahead of the HEMT amplifiers in the receivers and
sequentially changing the phase of each receiver local oscillator by
$90^{\circ}$. The correlated noise source is also used to measure
gain and phase variations in the instrument. Variations in the
calibration of the CBI, after applying the noise calibration
corrections, are at the 1\% and $1\deg$ level. The efficiency
of the receiving system is in the range 0.8--1.

The antennas have low-noise broadband HEMT amplifier receivers with
$\sim 25$~K noise temperatures. The typical system noise temperature
averaged over all ten bands is $\sim 30$ K, including ground spillover
and atmosphere.

The antennas can be moved from one location to
another on the platform to change the baseline lengths and
orientations.  The results presented here were made in two different
configurations; one (shown in Figure 1) with most of the antennas
around the perimeter of the platform, which provides a fairly uniform
$(u,v)$ coverage and allows easy access to the receivers; and a
second, more compact configuration which provides greater sensitivity
at low $l$.

One of the 13 receivers is currently configured for right circular
polarization while the other 12 receive left circular
polarization. The images presented here are from the
parallel-polarization baselines, which are sensitive to Stokes'
parameter $I$ (assuming circularly-polarized emission is
negligible). The 12 cross-polarized baselines are sensitive to Stokes'
parameters $Q\pm i U$ and are being used to place limits on the
polarization of the CMBR. The ability to rotate the array about the
optical axis facilitates polarization calibration and enables us to
measure both $Q$ and $U$ at the same $(u,v)$ point.

\section{The Site}

The CBI must be on a high, dry site to ensure that atmospheric
brightness fluctuations do not limit the sensitivity of the
instrument, and, in addition, the horizon must be low to minimize
pickup of ground radiation. We have chosen a site at an altitude of
5080 m about 40 km east of San Pedro de Atacama in northern Chile. The
site is near the center of the 10-km plateau of Llano de Chajnantor,
close to the site proposed for the Atacama Large Millimeter Array
(ALMA).
 
The CBI is housed in a retractable dome to provide some protection
from rain, snow and wind. On most days, the wind at Chajnantor peaks
late in the afternoon at about 15 m/s (30 mph), but we have
experienced winds greater than 45 m/s. Under these conditions wind
chill is a severe problem, but the dome provides a sheltered work
space where we can repair and maintain the CBI. The dome is a
hemispherical steel frame covered with polyester cloth, sitting atop a
2-m high wall 10.5 m in diameter.

The site facilities include a control room, laboratory, machine shop,
power plant, two bedrooms, and a bathroom -- all constructed within 
standard shipping containers.  All the equipment required for repairs and
modifications is at the site. Additional containers used for storage
are placed around the dome to provide a wind break.  To counteract the
effects of high altitude, the air in the work and living areas is
oxygen-enriched (using molecular sieves to separate oxygen from the
air), and people working outside can wear portable oxygen tanks with
demand regulators when necessary to improve efficiency and safety. The
power plant and fuel tanks are located about 100 m east of the CBI;
the average power consumption for the site is about 100 kW and the
power plant has a pair of diesel generators rated for 150 kW at 5000~m.

The base facilities are located in the historic oasis town of San
Pedro de Atacama at an elevation of 2500~m, and include bedrooms,
kitchen, conference room, and computer room; the facilities are
provided by the Hosteria La Casa de Don Tom{\'a}s.

\section{Project Status}

The CBI was designed in 1995--1996, and constructed in
1997--1998. After six months of test observations on the Caltech campus
in Pasadena, using six receivers and four 1-GHz correlators, the
instrument was partially disassembled, packed and shipped to Chile
where it arrived at the Llano de Chajnantor on August 28, 1999. We
obtained first light in Chile on November 1, 1999, and the full array
of 13 receivers and ten correlators was fully assembled and started
routine observations in January 2000.  Unfortunately the weather in
the first few months of 2000 has been unusually bad, and we have lost
about 50\% of the nights owing to cloud, high winds, and snow. In the
remaining nights the observing conditions were superb and fluctuations
in the atmosphere did not significantly increase the system noise. A
minor eruption of the nearby volcano Lascar on July 18 did not
interrupt observing.

The preliminary results presented here are from observations made
between January and July 2000. We will continue to make observations
in Chile until August 2001 or later.

\section{Observing Strategy}

For all likely CMBR power spectra, the expected signal decreases
strongly with interferometer baseline length (the expected
root-mean-square visibility is proportional to $C_l$), so we are
making observations in two modes: (1) deep observations of selected
45\arcmin\ fields, concentrating on the higher end of our $l$ range --
these observations will be limited by instrumental noise and
foreground emission; (2) shallower observations of many overlapping
fields (``mosaic'' observations) to measure the power spectrum at
lower $l$ -- in these observations we will image enough sky to balance
the effects of thermal noise and sample (cosmic) variance.  Our
observations have been made in the declination range $-5\deg <
\delta < -2\deg$ in regions of low IRAS 100~$\mu$m emission at several
galactic latitudes.

Daytime observations of the CMBR are not possible because radio
emission from the sun in the far sidelobes of the antennas
contaminates the visibilities; we also restrict observations to fields
more than $60\deg$ from the moon.

The dominant systematic contamination in the CBI observations is due
to radiation from the ground (``ground spillover''). This introduces
spurious signals up to a few Jy on the shortest (1-meter) baselines,
but the effect is much less on longer baselines. Fortunately the
ground signal is stable and repeatable on short time scales. We remove
it by differencing the visibilities measured on two fields observed at
the same elevation and azimuth. Usually we alternate between two
fields; we observe one ``leading'' field for 8~min, and then observe a
second ``trailing'' field 8~min later in right ascension for the next
8~min, so that both are observed at the same position relative to the
ground. By Fourier-transforming the visibility differences, we obtain
an image of the {\it difference} of two fields separated by 8~min in
right ascension. By comparing observations made at different
elevations and under different weather conditions, we have found that
any residual ground contamination in the difference images is less
than a few per cent of the CMBR signal at spatial frequencies
corresponding to the 1-meter baselines. While differencing requires
longer observing times than undifferenced observations, and can also
make it difficult to distinguish a positive $\delta T$ in one field
from a negative $\delta T$ in the other field, it does not otherwise
impede the measurement of the CMBR power spectrum.

The flux-density scale is based on observations of Jupiter, which has
been measured to have an effective temperature of $152\pm 5$~K at
32 GHz (Mason et al.\ 1999); as Jupiter does not have a simple thermal
spectrum between 26 and 36 GHz, we transfer this calibration to other
frequencies using observations of Tau~A (the Crab nebula) which has a
power-law spectrum in this range. Receiver gain and system-temperature
variations between observations of the calibration sources are removed
by reference to the internal calibration signal which is injected
into the front end of each receiver.

We make images from the visibility data by the usual synthesis-imaging
procedures (e.g., Taylor, Carilli, \& Perley 1999). The sensitivity of
an image made from a single pointing is tapered by the primary beam of
the antennas, approximately a gaussian of FWHM $45'\times(30\,{\rm
GHz}/\nu)$. In the mosaic observations, we typically observe at a grid
of positions spaced by $20'$ (e.g., 36 positions to map a
$2\deg\times2\deg$ field), and make a weighted linear combination
of the individual images (Cornwell 1988). The resulting mosaic image
has almost uniform sensitivity except at the edges.

The major foreground contaminant is discrete, unresolved extragalactic
sources (galaxies and quasars). In order to remove these sources from
the data, we have made 26--34 GHz observations of all the sources in
our fields that are stronger than 6 mJy in the 1.4 GHz NRAO VLA Sky
Survey (Condon et al.\ 1998), using the Owens Valley Radio
Observatory's 40-meter telescope. We monitor those that are stronger
than 6 mJy at 30 GHz using the 40-meter telescope to obtain their flux
densities at the time of the CBI observations. Then the expected
responses to these sources are subtracted from the CBI visibility data
before imaging.

\begin{figure}
\plottwo{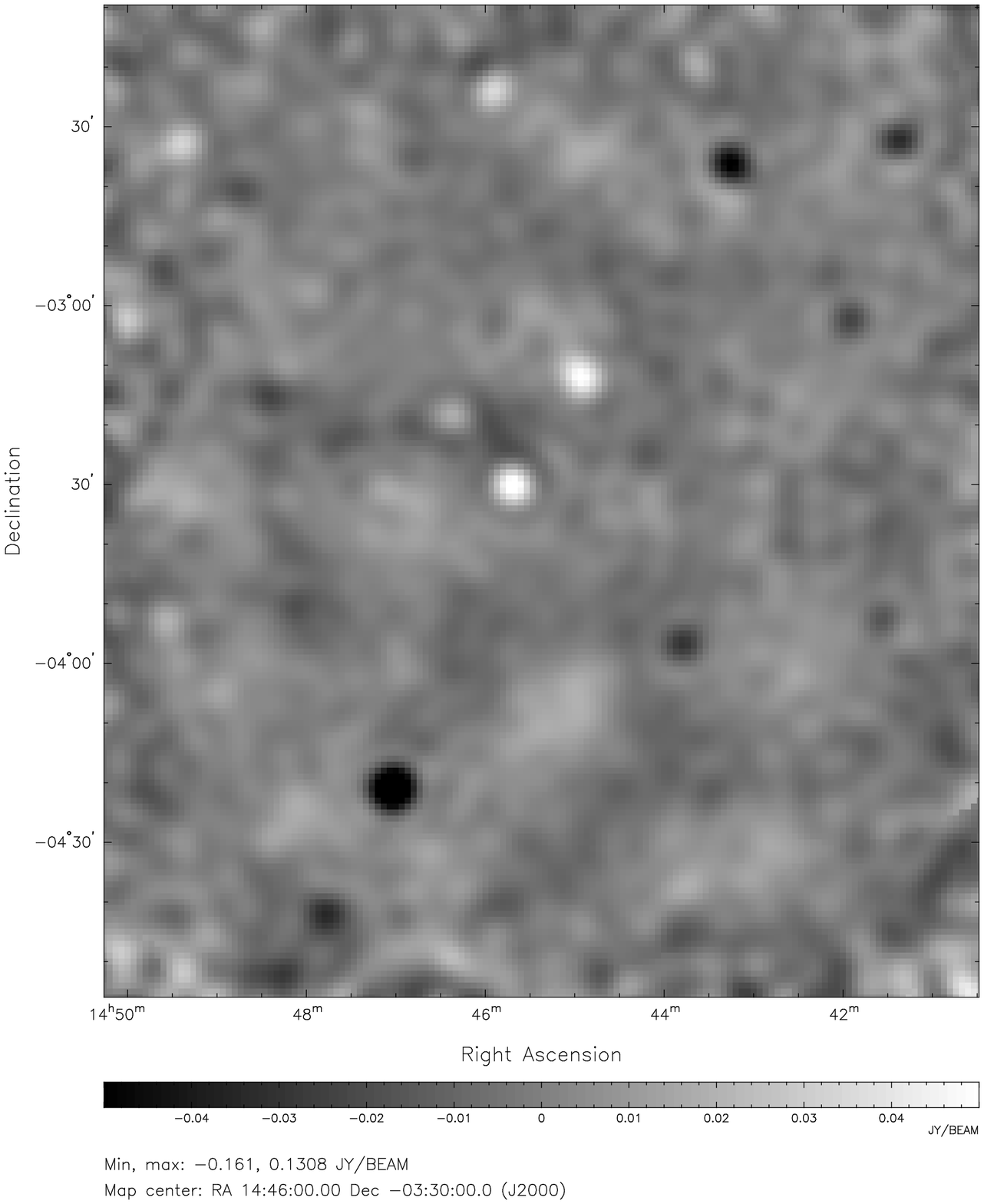}{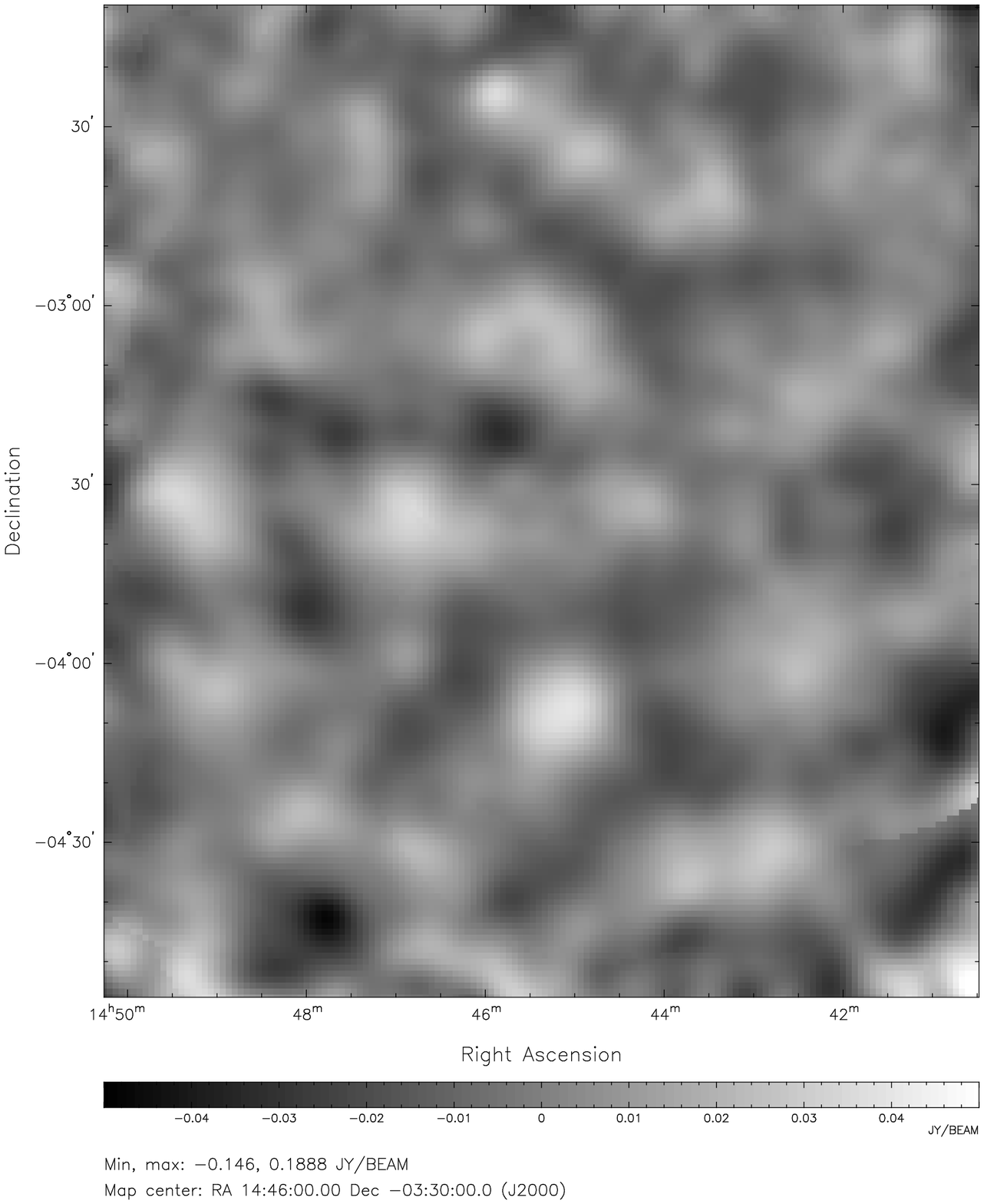}
\caption{Preliminary mosaic images of a $\sim2\deg$ field centered at
about RA $14^{\rm h}46^{\rm m}$, Decl. $-3\deg30'$. The images show
the {\it difference} of the emission in this field and one centered at
RA $14^{\rm h}54^{\rm m}$. Each field was observed in 50 single
pointings, with an integration time of about $2^{\rm h}$ on each
field.  {\it Left\/}: Before source subtraction; synthesized beam FWHM
5.2 arcmin. Light spots are discrete sources in the leading field,
while dark (negative) spots are discrete sources in the trailing
field. {\it Right\/}: After subtracting known sources using flux
densities measured with the OVRO 40-meter telescope; resolution
reduced to FWHM 7.3 arcmin to improve sensitivity to large-scale
emission. In this preliminary image, not all the known sources have
yet been subtracted.}
\end{figure}

Figure 2 shows one of our mosaic fields and the effectiveness of our
source subtraction technique.
The images are dominated by CMBR, and in future work we will quantify
this. There are a number of potential problems that we need to
understand before we are ready to determine the CMBR power spectrum,
however.  First, we must look for possible systematic errors due to
instrumental problems, atmosphere, or residual ground spillover. These
systematics are all largely removed by our differencing strategy, but
we can look for residual effects by comparing observations made at
different times, and on distinct but equal baselines. Such tests have
already shown that any contamination is less than a few percent of the
CMBR signal. Secondly, it is clear that we can accurately remove the
strongest discrete sources, but unsubtracted sources and errors in the
flux densities of the subtracted sources will bias the power spectrum.
We will measure the number counts of fainter sources and make a
statistical correction for them. Thirdly, we have yet to
determine how much of the emission seen in our images could be
foreground from the Galaxy. The ten CBI frequency bands allow us to
constrain the spectrum of the emission, and preliminary results show
that the spectrum is consistent with thermal emission. Synchrotron or
free-free foreground emission must be much less than the CMBR.

\section{Estimating the Power Spectrum}

We will use a maximum likelihood method to estimate the power spectrum.
If the CMBR fluctuations are gaussian, then the required information
about their power spectrum is contained in the covariance matrix
$M_{jk}$ of the visibility measurements. The covariance between a
visibility measurement $j$ at $(u,v)$ position ${\bf u}_j$ and frequency
$\nu_j$ and another measurement $k$ is related to the power spectrum
by
\begin{equation}
M_{jk} = \langle{V({\bf u}_j, \nu_j) V^*({\bf u}_k, \nu_k)}\rangle = K_j K_k \int_0^\infty W_{jk}(v) C(v) v dv ,
\end{equation}
where $K_j \equiv 2\nu_j^2 k_{\rm B} T_0 g(\nu_j)/c^2$ is the conversion
factor from $\delta T/T$ to intensity, $C(v) = C_l$ for $l\approx 2\pi
v$, and $W_{jk}(v)$ is the {\it window function} defined by
\begin{equation}
W_{jk}(v) = \int_0^{2\pi} \tilde A({\bf u}_j - {\bf v}, \nu_j) \tilde A^*({\bf u}_k - {\bf v}, \nu_k) d\theta_v,
\end{equation}
where $\tilde A$ is the normalized Fourier transform of the primary beam
response (Hobson, Lasenby, \& Jones 1995; White et al.\ 1999).
The diagonal window function, $W_{jj}(v)$, is approximately a gaussian
of FWHM 70 wavelengths ($\Delta l \approx 430$) for the CBI.
Using this expression we can compute the expected rms visibility for
a flat power spectrum 
\begin{equation}
l(l+1)C_l/2\pi = \left( \delta T/T_{\rm cmb}\right) ^2 = {\rm constant}.
\end{equation}
The result is
\begin{equation}
\sqrt{\langle{V^2}\rangle} \approx 1.1\, {\rm mJy} (d/{\rm m})^{-1} (\delta T/ \mu{\rm K}),
\end{equation}
where $d$ is the baseline length.
Typical cosmologies have $\delta T = 60$ $\mu$K for $l \sim 600$ and 
$\delta T = 15$ $\mu$K for $l \sim 2500$; the corresponding rms visibilities
are about $50$ mJy and $3$ mJy.

Additional terms must be included in the covariance matrix to account
for instrumental noise (which to a good approximation is uncorrelated
between different measurements and can be accurately measured) and
other errors including unsubtracted point sources and errors in
foreground subtraction.
Given the covariance matrix, we can calculate the likelihood of the
observed visibility data for a particular model power spectrum.

Observations of mosaiced fields will be used to estimate the power
spectrum with increased resolution in $l$. In a mosaic, the $l$
resolution is not limited by the inverse of the width of the primary
beam but improves in proportion to the size of the field surveyed. For
a field of size $2\deg$, the resolution in $l$ is about 160, and
will enable us to resolve the acoustic peaks.  Mosaicing also reduces
the cosmic variance as a larger area is being surveyed. The covariance
matrix for a mosaic observation is more complicated than that given
above, as it involves correlations between different pointing centers
as well as those between different baselines and frequencies, and the
dimension of the matrix is correspondingly increased, to $> 10^4
\times 10^4$ (number of pointings times number of distinct baselines
times number of frequency channels).

\section{Conclusions}

The preliminary results presented here show that the CBI is working
well. We will soon submit a paper to the {\it Astrophysical Journal}
presenting the first quantitative results of our observations on the
intrinsic anisotropy of the CMBR at $l < 1500$ (Padin et al.\
2000c). In subsequent papers we will extend the results to higher $l$,
which will require careful attention to the effects of discrete
sources, and use mosaic observations to improve the resolution in $l$.
While so far we have concentrated on ``blank'' regions for measurement
of the intrinsic power spectrum, we are also observing a sample of
galaxy clusters to study the Sunyaev-Zeldovich effect (Udomprasert et
al.\ 2000).

\acknowledgments We are grateful for the contributions to this
project of our collaborators: Russ Keeney, Steve Miller, Walter
Schaal, and John Yamasaki (Caltech); John Carlstrom and Erik Leitch
(University of Chicago); Bill Holzapfel (University of California,
Berkeley); Steven Myers (National Radio Astronomy Observatory);
Marshall Joy (NASA's Marshall Space Flight Center); Angel Otarola
(European Southern Observatory); and Leonardo Bronfman, Jorge May,
Simon Casassus, and Pablo Altamirano (University of Chile).
The CBI project has been supported by the National Science Foundation
under grants AST-9413935 and AST-9802989, and we are also grateful for
the generous support of Maxine and Ronald Linde, Cecil and Sally
Drinkward, and our colleagues at the California Institute of
Technology, especially the Provost, the President, and the Chairman of
the Division of Physics, Mathematics, and Astronomy.  We are grateful
to CONICYT for permission to operate the CBI in the Chajnantor
Scientific Preserve in Chile. JS and PSU acknowledge support from
National Science Foundation Graduate Student Fellowships.

\end{document}